\pdfoutput=1

\documentclass[a4paper]{jpconf}

\usepackage{graphicx}
\graphicspath{{fig/}}

\usepackage{hyperref}
\usepackage{subcaption}
\usepackage{amsmath,amssymb}

\newcommand{\tHq}{\ensuremath{tHq}}
\newcommand{\ttbar}{\ensuremath{t\bar t}}

\hypersetup{pdfauthor={Andrey Popov on behalf of the CMS collaboration}, %
 pdftitle={Identification of signal events in a search for H->bb produced in association with single top quarks}}

\begin{document}

\title{Identification of signal events in a search for \boldmath$H\to b\bar b$ produced in association with single top quarks}

\author{Andrey Popov on behalf of the CMS collaboration}

\address{Universit\'{e} catholique de Louvain, Louvain-la-Neuve, Belgium; also at Lomonosov Moscow state university, Moscow, Russia}

\ead{andrey.popov@cern.ch}

\begin{abstract}
The CMS collaboration has performed a search for an associated production of Higgs bosons and single top quarks in the $H\to b\bar b$ decay channel.
The measurement is challenged by a complex multijet final state and an overwhelming background from top-quark pair production.
These proceedings address a particular aspect of the search and describe the procedure deployed to discriminate the signal process from backgrounds.
\end{abstract}

\section{Introduction}

After the ATLAS and CMS collaborations reported a discovery of a new particle~\cite{ATLASHiggsDiscovery, CMSHiggsDiscovery} that resembles a Higgs boson, its properties are under investigation.
Measurements of couplings of the particle conducted so far~\cite{ATLAS-CONF-2014-009, CMS-PAS-HIG-14-009} agree with expectations from the standard model (SM).
However, if one allows contributions from possible new physics at the loop level, the constraints on coupling constants relax and a second allowed region emerges around the point $\kappa_f = -1$, $\kappa_V = +1$~\cite{EllisYou13}, where $\kappa$ denotes a scale factor of the coupling constants to fermions or vector bosons with respect to SM values and the relative sign between the two scale factors is physical.

\begin{figure}[hbt]
\begin{center}
 \subcaptionbox{\label{Fig:Diagrams:tHq}}{\includegraphics[]{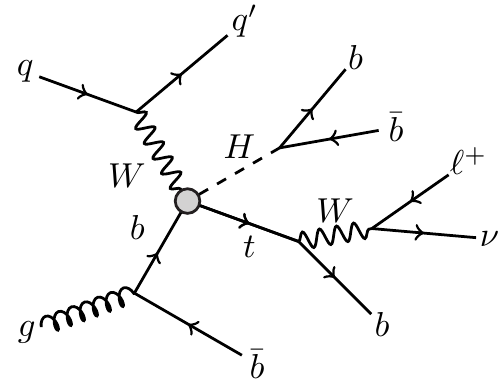}}\qquad
 \subcaptionbox{\label{Fig:Diagrams:ttbar}}{\includegraphics[]{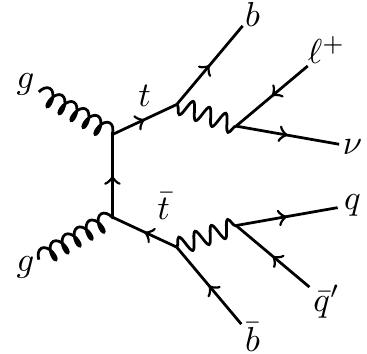}}
\end{center}
\caption{Representative Feynman diagrams with full decay chains for signal process~(a) and semileptonic~\ttbar{}~(b), which is the dominant background.}
\label{Fig:Diagrams}
\end{figure}

Associated production of Higgs bosons and single top quarks provides a unique handle to discriminate between the two allowed regions~\cite{Biswas13, Farina13}.
A representative Feynman diagram for this process (hereafter referred to as \tHq{}) is shown in figure~\ref{Fig:Diagrams:tHq}.
The Higgs boson can couple to either the top quark or the $t$-channel $W$~boson. In SM the two types of diagrams interfere destructively and nearly compensate each other~\cite{MaltoniPaul01}.
However, the interference is constructive if $\kappa_f = -1$, and the production cross section increases by more that an order of magnitude~\cite{Biswas13, Farina13}.

Searches for the \tHq{} production with $\kappa_f = -1$ have been performed in the CMS experiment~\cite{CMS} in the $H\to\gamma\gamma$~\cite{CMS-PAS-HIG-14-001} and $H\to b\bar b$~\cite{CMS-PAS-HIG-14-015} decay channels.
These proceedings are focused on the procedure adopted in the $H\to b\bar b$ channel to identify signal events.
Other aspects of this search are discussed in detail in reference~\cite{CMS-PAS-HIG-14-015}.

\section{Event selection}

The top quark is required to decay semileptonically, producing a muon or an electron, while for the Higgs boson the decay $H\to b\bar b$ with the largest branching ratio is considered.
The three $b$-quark jets from the decays are accompanied by a forth one (see figure~\ref{Fig:Diagrams:tHq}), which, however, sometimes falls outside the tracker acceptance and then cannot be identified as originating from a $b$~quark, or $b$-tagged.
Consequently, events must contain three or four $b$-tagged jets.
Another prominent feature of the signal process is the light-flavour recoil jet (denoted as $q'$ in figure~\ref{Fig:Diagrams:tHq}), which is often forward.
The event selection therefore requires the presence of at least one untagged jet, accounting for the recoil jet and possible additional radiation.

The dominant background stems from semileptonic decays of \ttbar{}.
Because of the small \tHq{} production cross section the signal-to-background ratio in the $\kappa_f = -1$ case is about 0.7\% (2\%) in the region with three (four) $b$-tagged jets.
This makes it essential to deploy methods of multivariate analysis (MVA).
The measurement exploits artificial neural networks trained with the Broyden--Fletcher--Goldfarb--Shanno (BFGS) algorithm as implemented in the TMVA package~\cite{TMVA}.

\section{Jet assignment under the \texorpdfstring{\boldmath$tHq$}{tHq} hypothesis}

The multijet final state poses a challenge to construct observables that discriminate between signal and background processes.
In order to do it efficiently, it is important to hypothesise the origin of all reconstructed jets in an event.
This problem can be addressed by a kinematic fit, imposing constraints from masses of the Higgs boson and top quark.
However, the measurement adopts a different approach, which makes use of not only momenta but also detector-related information such as $b$-tagging decisions.

The goal of the procedure is to identify reconstructed jets stemming from the four quarks in the final state $tHq \to l\nu 3bq$.
For each \tHq{} event all possible ways to assign reconstructed jets to the four quarks are considered.
The three $b$~quarks are almost always central, and this is used to simplify the combinatorics.
Each particular way to choose four jets and assign them to the four quarks represents a possible interpretation of the event.
An interpretation is said to be correct if four-momentum of each of the four jets is close, in the $\Delta R = \sqrt{\Delta\eta^2 + \Delta\phi^2}$ metrics, to the four-momentum of the quark to which the jet is assigned; otherwise the interpretation is declared wrong.

Each interpretation can be described by a number of observables, which can be chosen to discriminate between correct and wrong interpretations.
A dedicated MVA is trained to perform the discrimination using these observables as input variables.
The list of variables includes masses of the reconstructed Higgs boson and top quark, angular separations between various decay products, $b$-tagging information, and other quantities.

In order to perform the jet assignment in an unknown event, under an assumption that it stems from the signal process, all possible event interpretations are constructed, the trained MVA is evaluated for each one, and the interpretation with the largest MVA response is accepted.
Among \tHq{} events that have correct interpretations, the jet from the decay $t\to bl\nu$, both jets from $H\to b\bar b$, and the recoil jet are identified correctly in approximately 66\%, 65\%, and 79\% of cases respectively.

\section{Jet assignment under the semileptonic \texorpdfstring{\boldmath$t\bar t$}{ttbar} hypothesis}

Top-quark pair production with semileptonic decays is by far the dominant background.
In order to provide an additional handle to suppress it, a dedicated jet assignment under the \ttbar{} hypothesis is performed.
Identifying origins of jets in a \ttbar{}~event allows to define in a natural way observables that are intrinsic to this process such as the mass of the hadronically decaying top quark.

Same approach as described in the previous section is followed.
All possible ways to assign reconstructed jets to the four quarks in the final state shown in figure~\ref{Fig:Diagrams:ttbar} are considered.
To limit the number of event interpretations, only central $b$-tagged jets are assigned to the two $b$~quarks in the final state.
An MVA is trained to distinguish correct and wrong event interpretations using similar classes of input variables.
The $b$-quark jets from $t\to bl\nu$ and $t\to\text{hadrons}$, and both non-$b$-quark jets from $t\to\text{hadrons}$ are identified correctly in about 66\%, 68\%, and 57\% of semileptonic~\ttbar{} events that have correct interpretations.

\section{Discrimination between the signal and backgrounds}

A third MVA is deployed to discriminate between the signal process and a mixture of semileptonic and dileptonic~\ttbar{} and $t\bar tH$.
The jet assignment described above is only exploited to construct input variables for this MVA.
Each event surviving the selection is interpreted under the both hypotheses in parallel, which allows to define two sets of observables.

\begin{figure}[hbt]
\begin{center}
 \subcaptionbox{}{\includegraphics[width=0.24\textwidth]{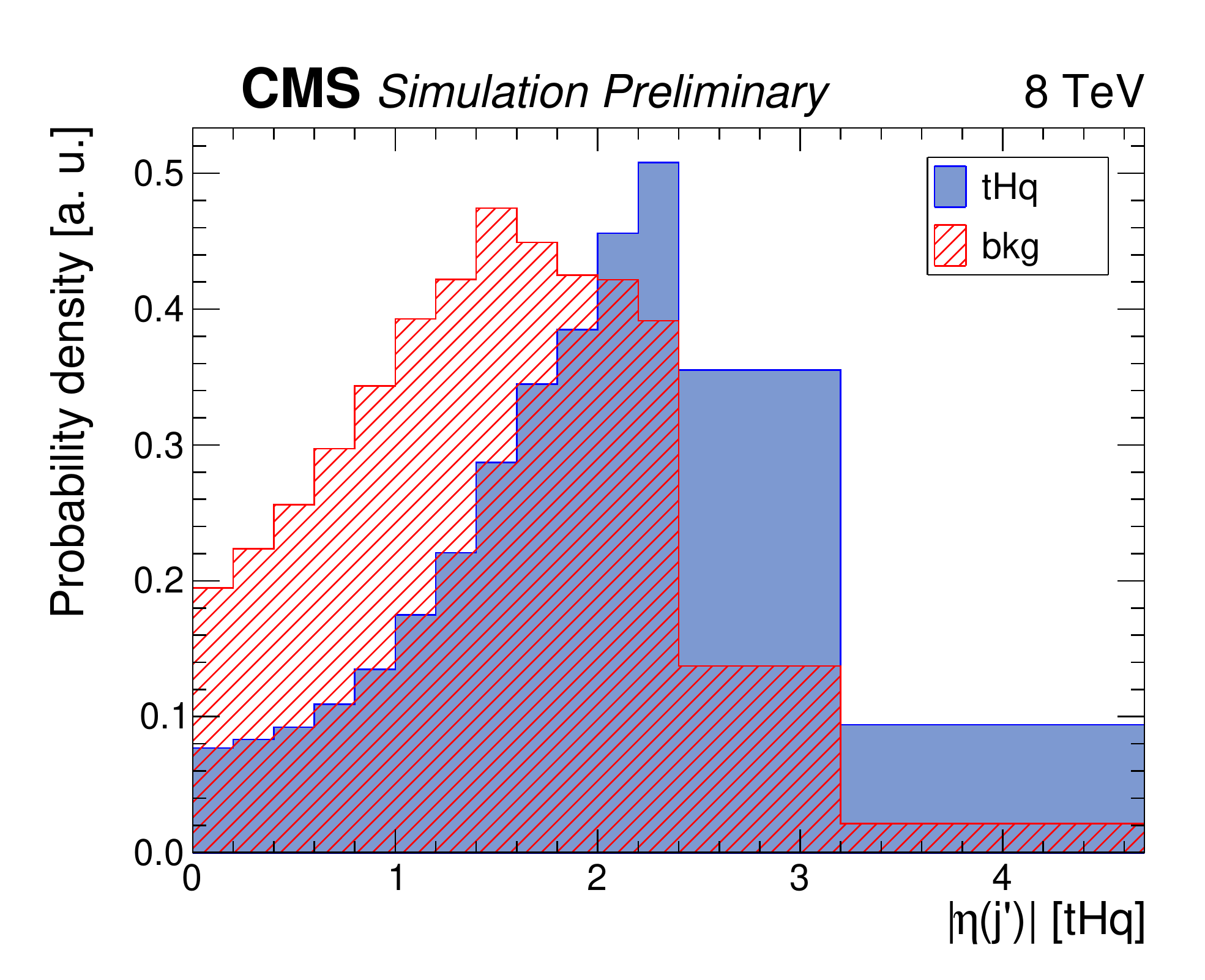}} %
 \subcaptionbox{}{\includegraphics[width=0.24\textwidth]{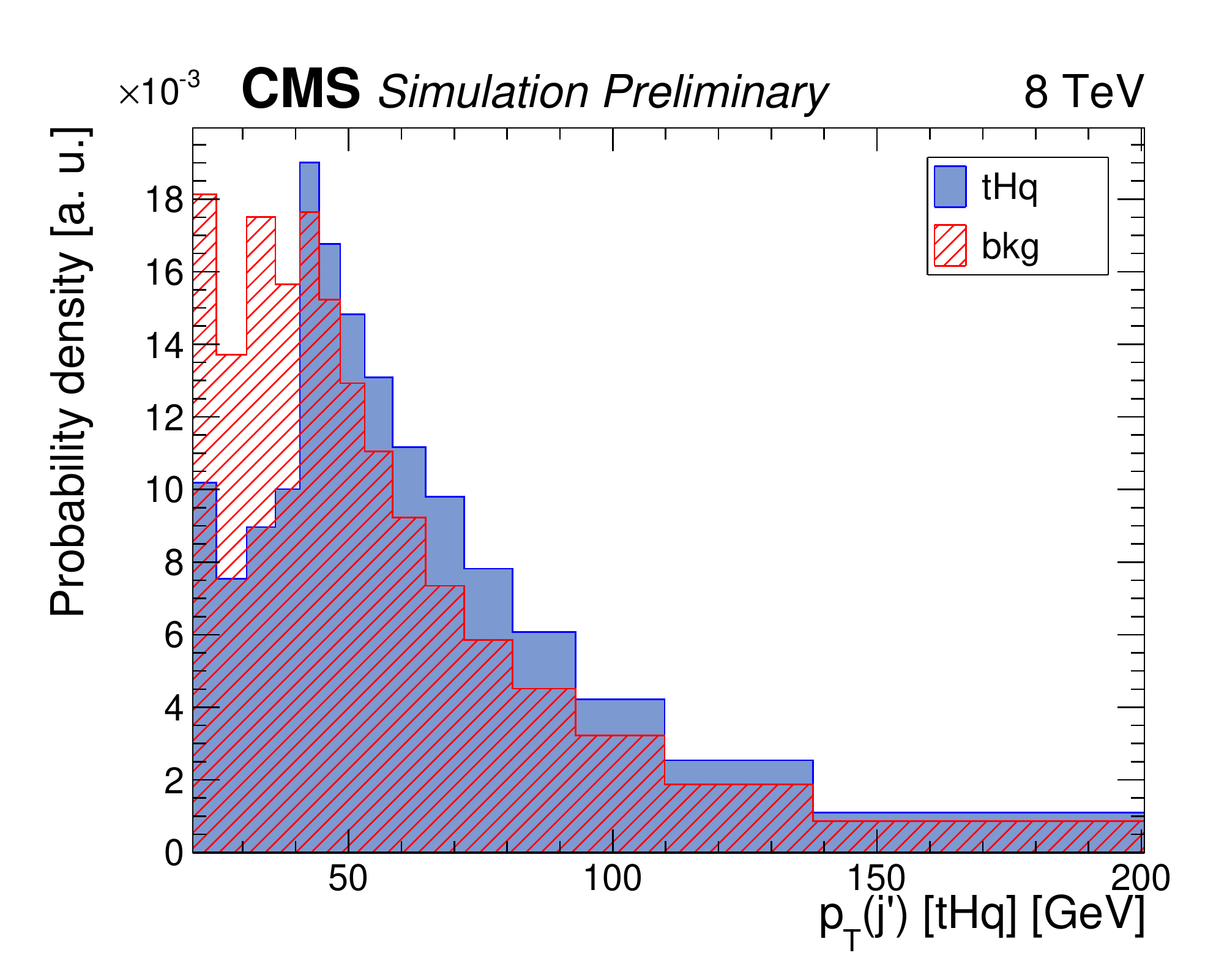}} %
 \subcaptionbox{}{\includegraphics[width=0.24\textwidth]{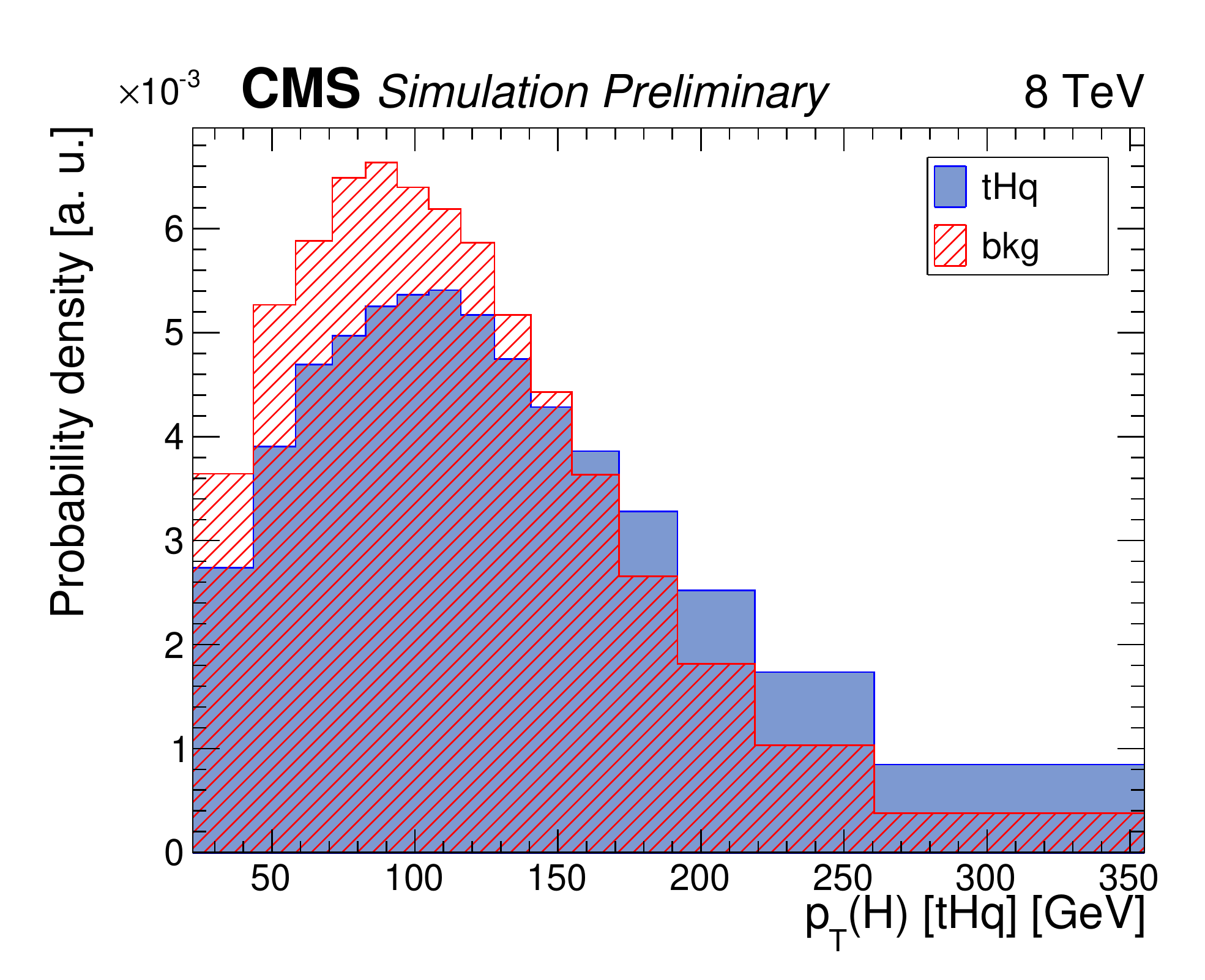}} %
 \subcaptionbox{}{\includegraphics[width=0.24\textwidth]{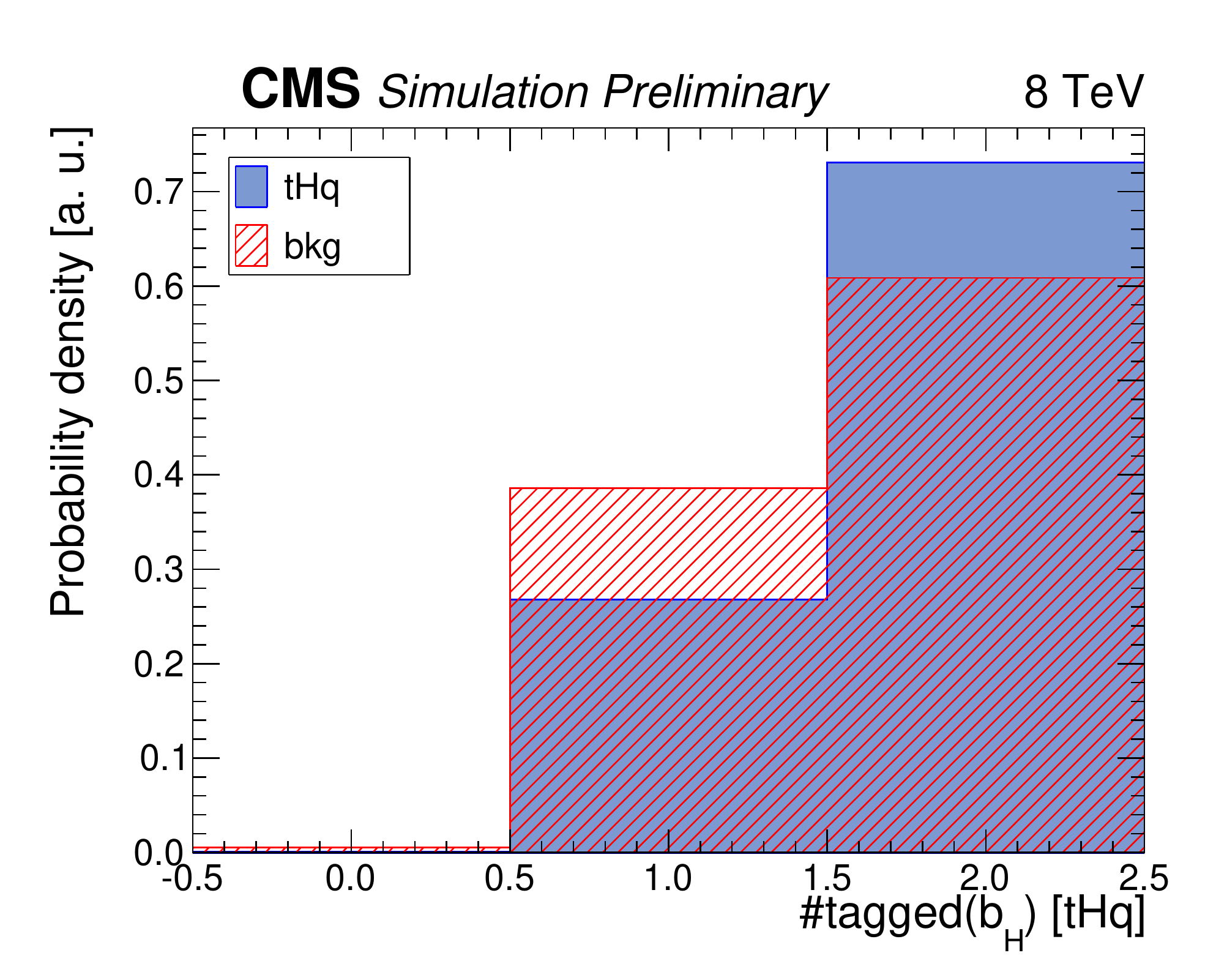}}
 
 \subcaptionbox{}{\includegraphics[width=0.24\textwidth]{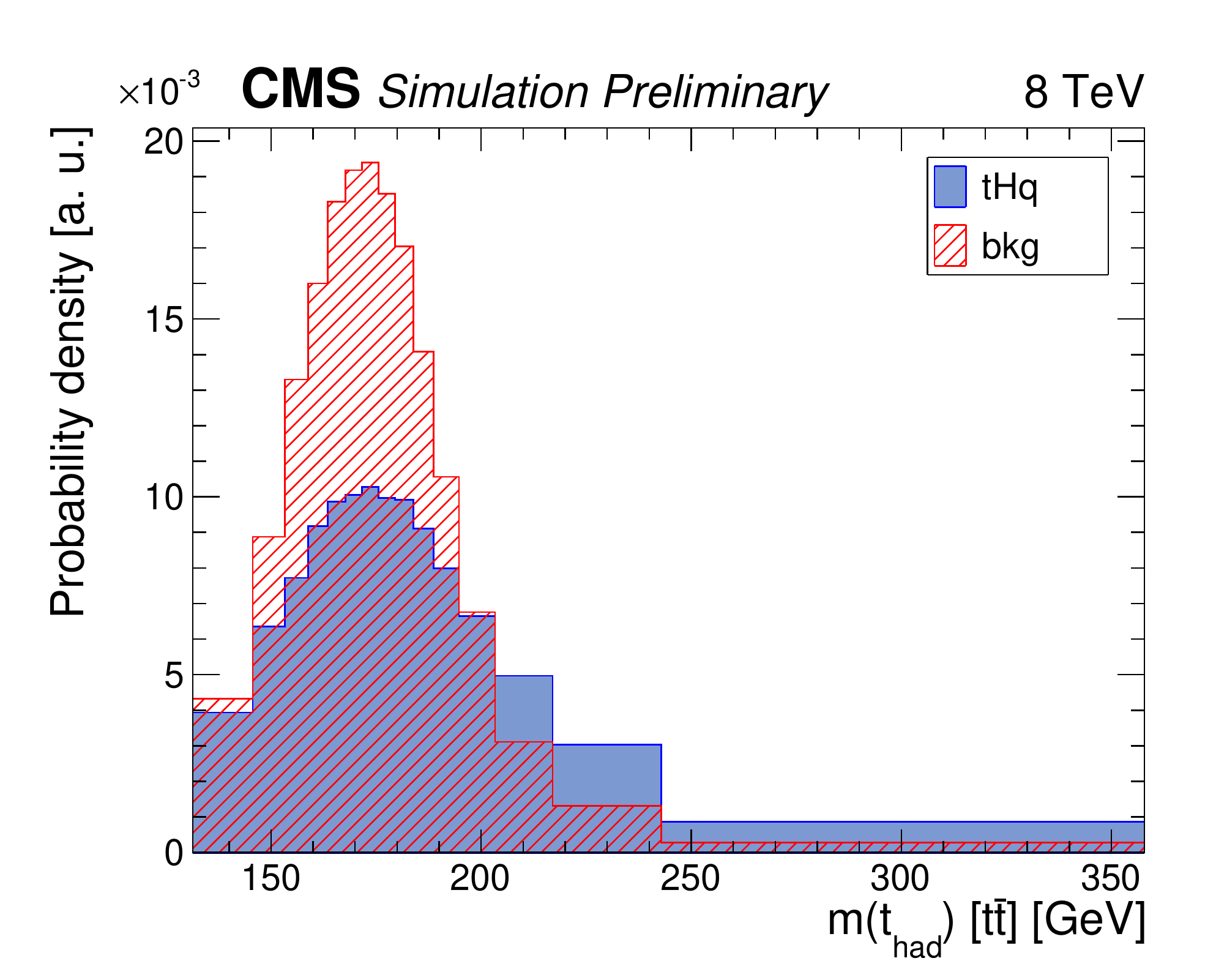}} %
 \subcaptionbox{}{\includegraphics[width=0.24\textwidth]{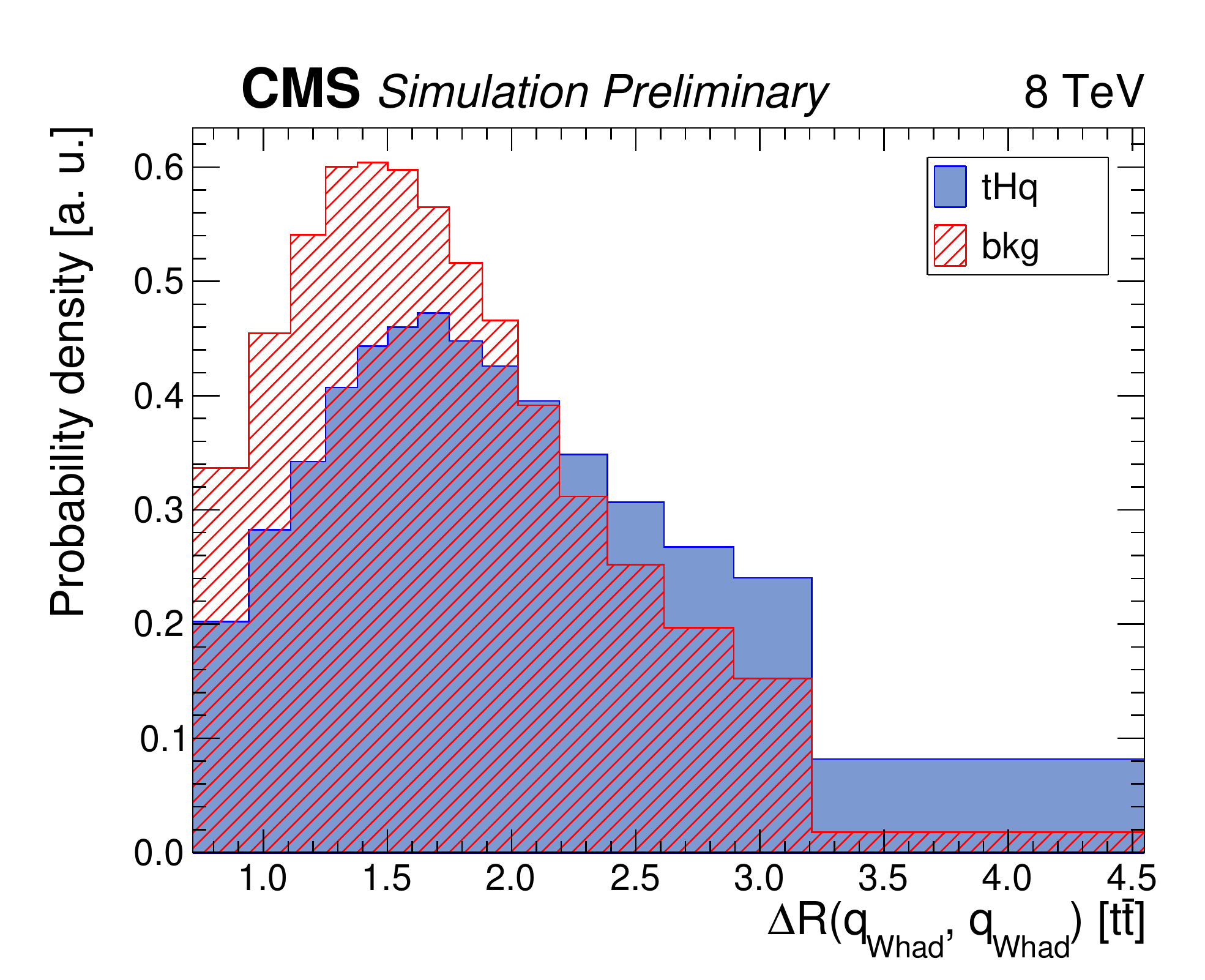}} %
 \subcaptionbox{}{\includegraphics[width=0.24\textwidth]{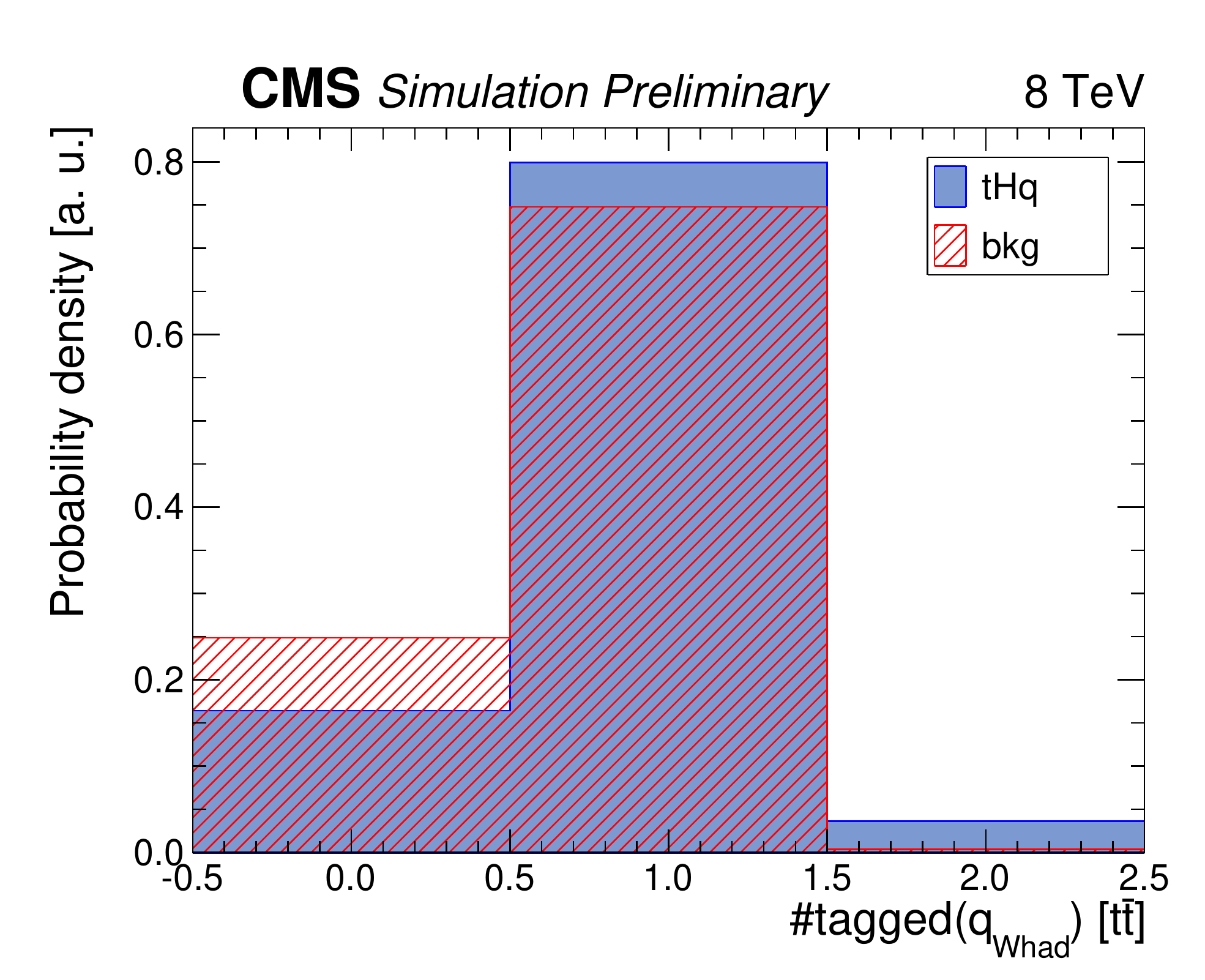}} %
 \subcaptionbox{}{\includegraphics[width=0.24\textwidth]{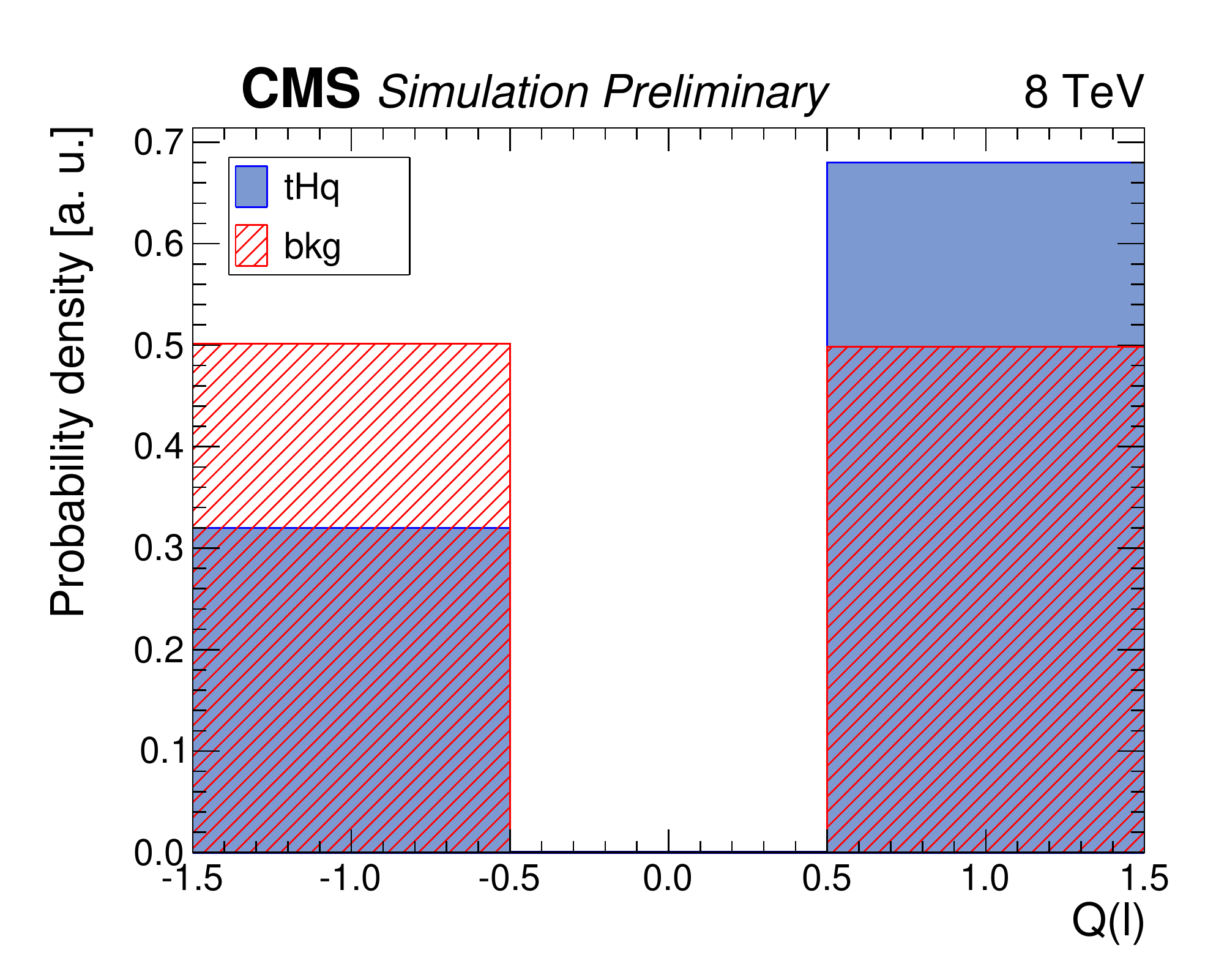}}
\end{center}
\caption{Distributions of signal and background events in input variables used in the classification MVA.}
\label{Fig:ClassInputVars}
\end{figure}

A relatively large set of $\mathcal O(20)$ input variables was considered as a starting point.
It was then optimised by excluding variables recursively, one at a time, until a significant degradation in the performance of the MVA was observed.
The last set before the drop in performance is chosen.
It consists of eight variables, and their distributions for signal and background events are shown in figure~\ref{Fig:ClassInputVars}.
Four variables are constructed under the \tHq{}~hypothesis: pseudorapidity and $p_\text{T}$ of the recoil jet~(a) and (b), $p_\text{T}$ of reconstructed Higgs boson~(c), and number of $b$-tagged jets assigned to the Higgs boson~(d).
Another three observables assume the \ttbar{}~hypothesis: mass of the hadronically decaying top quark~(e), $\Delta R$ between the two non-$b$-quark jets in $t\to \text{hadrons}$~(f), and number of $b$-tagged jets among the two~(g).
The last input variable is the electric charge of the lepton~(h), and it does not depend on jet assignment.

\begin{figure}[htb]
\begin{center}
 \subcaptionbox{}{\includegraphics[width=0.4\textwidth]{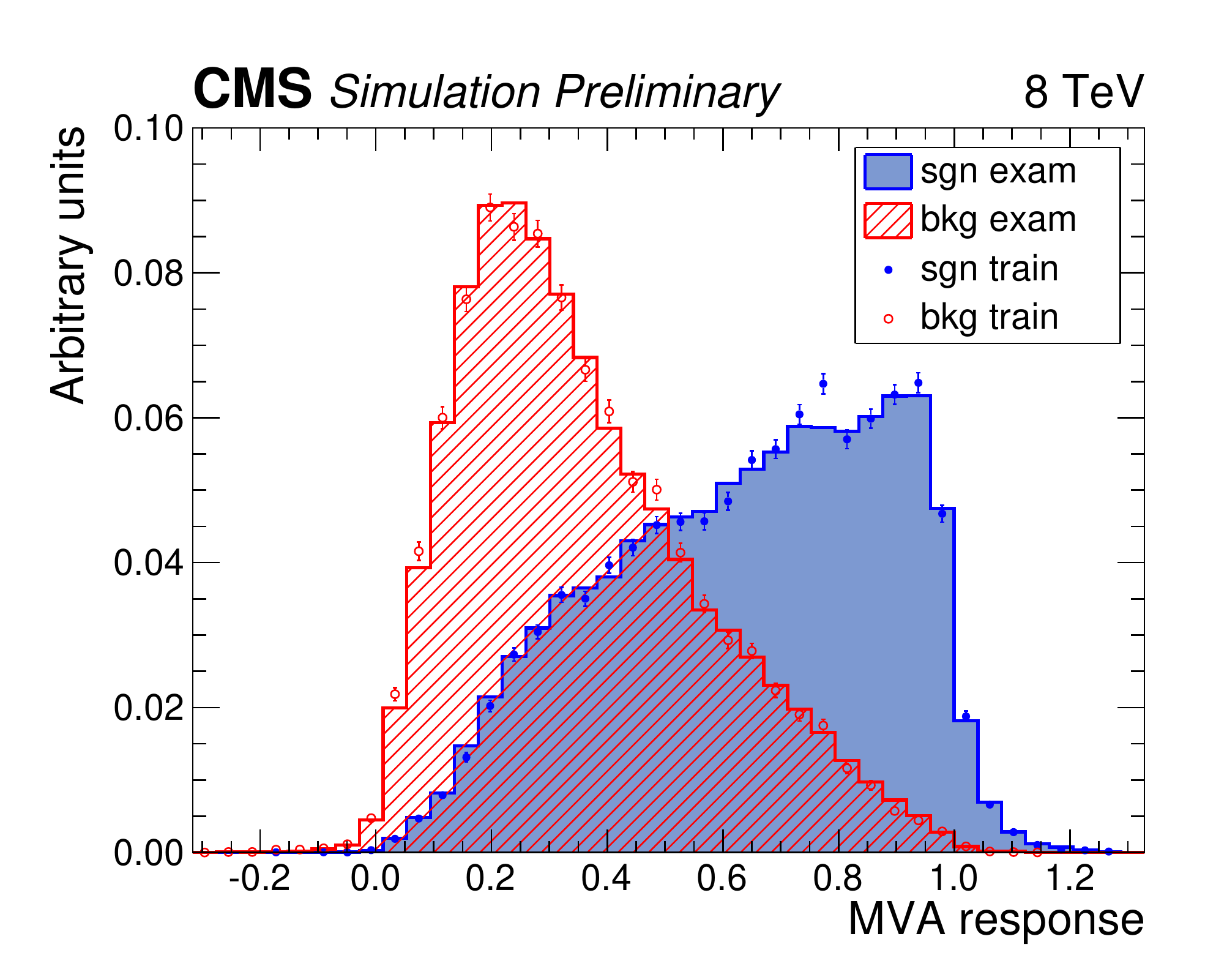}} %
 \subcaptionbox{}{\includegraphics[width=0.4\textwidth]{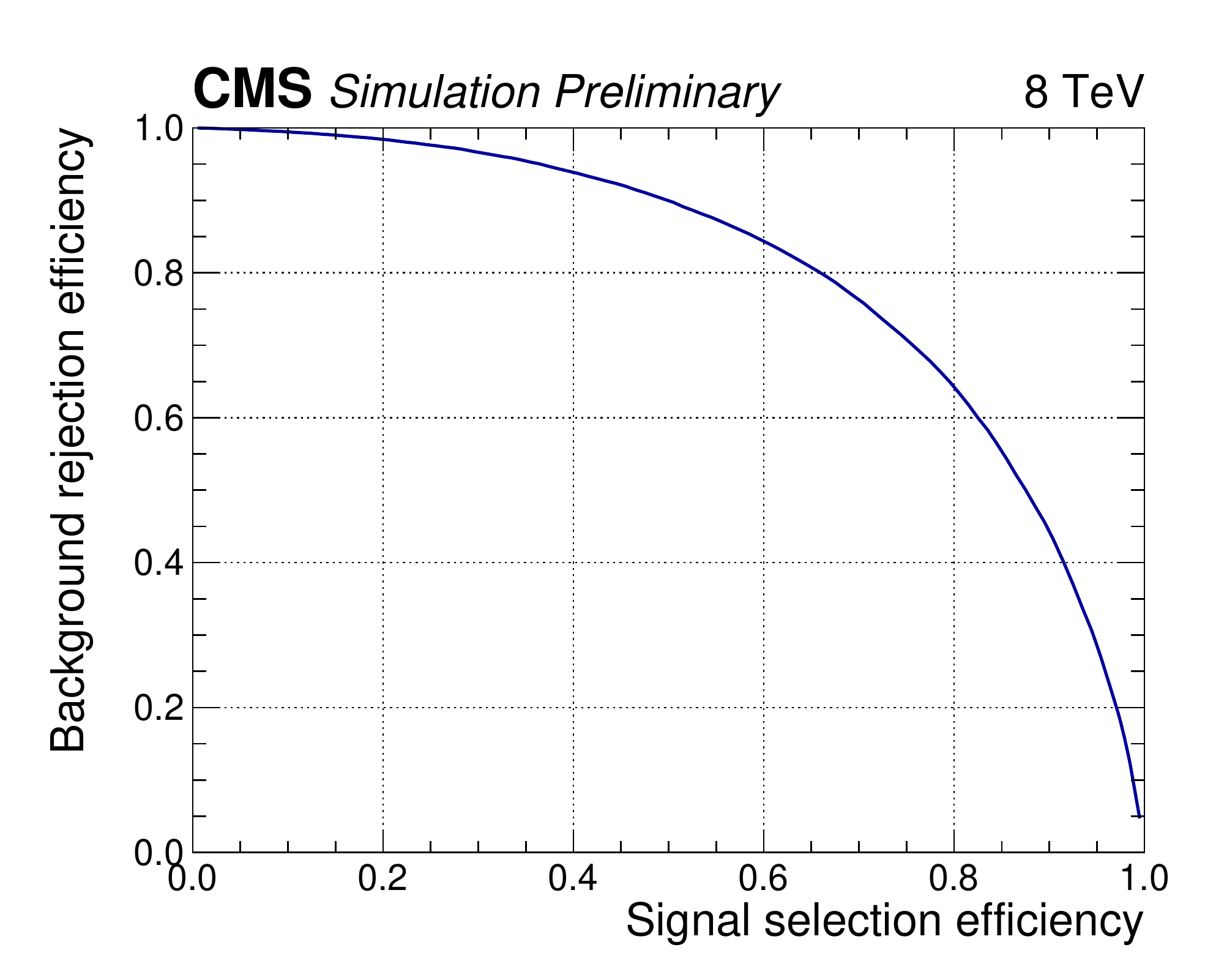}}
\end{center}
\caption{Distribution of signal and background events in response of the classification MVA~(a) and its ROC curve~(b).}
\label{Fig:ClassPerformance}
\end{figure}

Distribution of the response of the event classification MVA and its performance are shown in figure~\ref{Fig:ClassPerformance}.
The signal cross section is extracted by fitting the response in data.

\section{Summary}

Identification of signal events in a search for the \tHq{} production in the $H\to b\bar b$ decay channel was discussed.
Methods of multivariate analysis were exploited because of a small signal-to-background ratio.
Input variables were defined with the help of an advanced procedure of event reconstruction.
The search excluded a production cross section larger than 1.8\,pb at a 95\% confidence level~\cite{CMS-PAS-HIG-14-015}.



\end{document}